\documentclass[twocolumn,times]{aastex63}

\usepackage{CJKutf8}
\usepackage{amsmath}
\usepackage{amssymb}
\usepackage{amsfonts}
\usepackage{graphicx}
\usepackage{wasysym}
\usepackage{multirow}
\usepackage{mdframed}
\usepackage[T1]{fontenc}

\newcommand{\mjypbm}{\mbox{mJy\,beam$^{-1}$}}

\newcommand{\msun}{\mbox{\,$M_\odot$}}

\hypersetup{linkcolor=green,citecolor=blue,filecolor=cyan,urlcolor=magenta}
\hypersetup{linkcolor=red,citecolor=blue,filecolor=cyan,urlcolor=magenta}

\received{}
\revised{}
\accepted{}
\submitjournal{ApJ}

\shorttitle{}
\shortauthors{Cheng et al.}

\begin{document}
\begin{CJK}{UTF8}{gbsn}

\title{Low-Metallicity Star Formation Survey in Sh2-284 (LZ-STAR): The Core Mass Function}

\correspondingauthor{Yu Cheng}
\email{ycheng.astro@gmail.com}

\author[0000-0002-0786-7307]{Yu Cheng (程宇)}
\affil{National Astronomical Observatory of Japan, 2-21-1 Osawa, Mitaka, Tokyo 181-8588, Japan}

\author[0000-0002-3389-9142]{Jonathan C. Tan} 
\affiliation{Dept. of Astronomy, University of Virginia, Charlottesville, Virginia 22904, USA}
\affiliation{Department of Physics and Astronomy, Chalmers University of Technology, 412 93 Gothenburg, Sweden}

\author[0000-0002-5306-4089]{Morten Andersen}
\affiliation{European Southern Observatory, Karl Schwarzschild Str. 2, 85748, Garching, Germany}

\author[0009-0008-6570-9287]{Alva Kinman}
\affiliation{Department of Physics and Astronomy, Chalmers University of Technology, 412 93 Gothenburg, Sweden}

\author[0000-0003-4040-4934]{Rub\'{e}n Fedriani}
\affiliation{Instituto de Astrof\'isica de Andaluc\'ia, CSIC, Glorieta de la Astronom\'ia s/n, E-18008 Granada, Spain}

\author[0000-0002-6907-0926]{Kei E. I. Tanaka}
\affiliation{Department of Earth and Planetary Sciences, Institute of Science Tokyo, Meguro, Tokyo, 152-8551, Japan}

\author[0000-0001-7511-0034]{Yichen Zhang}
\affiliation{Department of Astronomy, School of Physics and Astronomy, Shanghai Jiao Tong University, 800 Dongchuan Road, Shanghai 200240, People's Republic of China}
\affiliation{State Key Laboratory of Dark Matter Physics, School of Physics and Astronomy, Shanghai Jiao Tong University, Shanghai 200240, People's Republic of China}
\affiliation{Key Laboratory for Particle Astrophysics and Cosmology (MOE)/Shanghai Key Laboratory for Particle Physics and Cosmology, Shanghai 200240, People's Republic of China}

\begin{abstract}

We present an ALMA 1.3~mm dust continuum study of the dense core mass function (CMF) in Sh2-284, a low metallicity outer Galaxy star-forming complex with $Z\sim1/3$--$1/2~Z_\odot$. The observations cover six far-infrared bright subregions at $\sim$0\farcs{65} (3000~au) resolution. We identify a total of 91 candidate dense cores and define a robust catalog of 68 cores with Gaussian fitting. The high-mass CMF above $2.5~\msun$ is well described by a Salpeter-like slope, with a fiducial forward-modeled value of $\alpha=1.22^{+0.24}_{-0.22}$ for $dN/d\log M\propto M^{-\alpha}$. Together with existing constraints on the initial mass function (IMF) of Sh2-284, the Salpeter-like CMF is consistent with a resemblance between the CMF and IMF shapes in the outer Galaxy environments, suggesting that moderately low metallicity alone does not strongly reshape the high-mass CMF/IMF slope.

\end{abstract}

\section{Introduction}\label{sec:intro}

The stellar initial mass function (IMF) is a fundamental distribution that links star formation and galaxy evolution. Observations of the Galactic field and nearby young clusters have revealed broadly similar IMFs, characterized by a Salpeter-like power-law slope above $\sim$1~\msun{} and a turnover near $\sim$0.2--0.3~\msun\   \citep{Kroupa01,Chabrier03,Bastian10}, yet variations have been reported in other Galactic and extragalactic environments \citep[e.g.,][]{Schneider18,Hosek19}. 
Metallicity is a key parameter in this context, as low-metallicity conditions are common in the early universe. Spatially resolved studies of nearby sub-solar-metallicity regions do not show a simple metallicity-dependent IMF trend. Several Magellanic Cloud associations appear broadly consistent with Galactic IMFs \citep[e.g.,][]{Schmalzl08,DaRio09,Andersen09}, whereas the massive-star IMF over 15--200~\msun{} is shallower than Salpeter in the 30~Doradus \citep{Schneider18}, and recent JWST studies of the outer Galaxy clusters suggest a lower turnover mass than the field IMF \citep{Yasui24,Andersen25}. 

The IMF is commonly hypothesized to be at least partly inherited from the core mass function (CMF), the mass distribution of dense cores before they form stars \citep{Offner14}. In solar neighborhood clouds, the CMF resembles the IMF and is shifted toward higher masses \citep[e.g.,][]{Alves07}. Over the past decade, extensive measurements with ALMA have been made in diverse Galactic environments, revealing high mass slopes ranging from a shallower index of $\sim$1 to the Salpeter-like index of $\sim$1.35 \citep[e.g.,][]{Motte18}. {Recent ALMA-IMF and ALMAGAL studies indicate that the CMF is not static, showing systematic differences between prestellar and protostellar core populations and shifting toward higher masses and shallower slopes as clumps evolve, favoring a clump-fed scenario in which cores continue to accrete material from their surroundings \citep[e.g.,][]{Nony23,Armante24,Coletta25}.} However, the CMF measurements in low-metallicity regions remain rare \citep[e.g.,][]{Traficante26}. While it is now clear that interpreting the CMF--IMF connection also requires accounting for processses such as fragmentation and core growth, observations at earlier stages of stellar mass assembly remain essential for testing whether possible IMF variations are already imprinted during cloud fragmentation, where metallicity may have a more direct influence.

Here we present an ALMA 1.3~mm dust-continuum measurement of the CMF in the outer Galaxy region Sh2-284. As one of the most metal-poor star-forming environments known in the Milky Way \citep[1/3--1/2 solar,][]{Negueruela15}, Sh2-284 is located at a favorable distance of 4.5~kpc \citep{Negueruela15}, providing a complementary low-metallicity laboratory in a less extreme environment than the strongly irradiated, starburst-like regions studied in the Large Magellanic Cloud, while being a factor of 10 closer. Building on the multi-wavelength LZ-STAR survey of the region \citep{Cheng25,Andersen25}, our ALMA data probe dense cores on $\sim$3000~au scales across multiple subregions, enabling a first dust-continuum-based CMF measurement in the low metallicity outer Galaxy environment.

\begin{figure*}[ht!]
\epsscale{1.1}\plotone{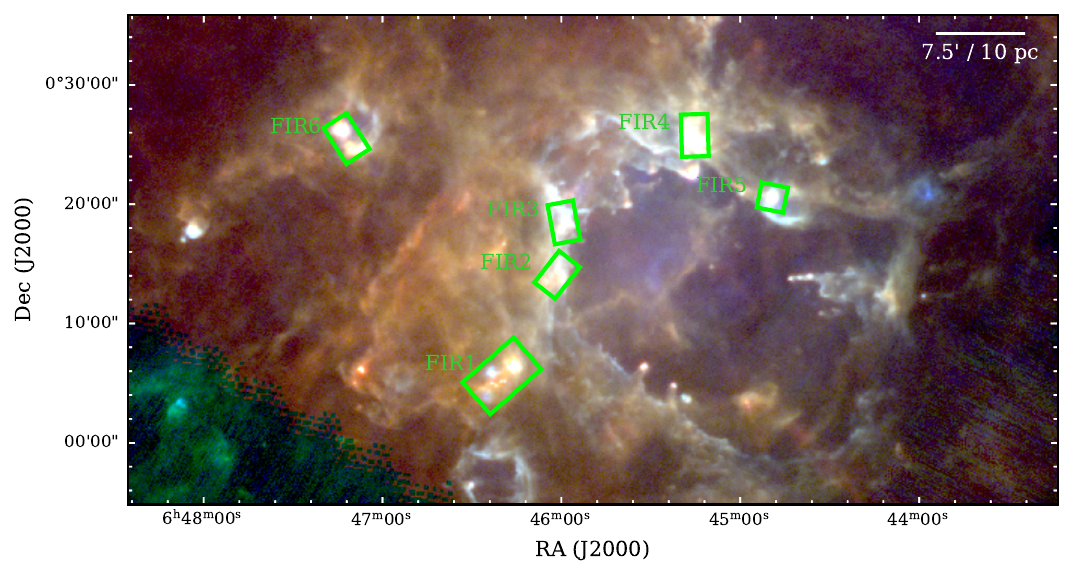}
\caption{Color composite image of the Sh2-284 region constructed from {\it Herschel} {70~$\mu$m (blue), 160~$\mu$m (green), and 500~$\mu$m (red)} images. The green rectangles show the ALMA mosaic fields. 
}\label{fig:overview}
\end{figure*}

\section{Observations}\label{sec:obs}

The six targeted regions (FIR1 to FIR6) in Sh2-284 are shown in \autoref{fig:overview} and summarized in \autoref{table:sum}. They represent the most active star-forming sites in the region, as indicated by strong dust emission in the {\it Herschel} images and numerous young stars identified in infrared surveys \citep{Puga09}. The ALMA {band~6} observations were carried out in Cycle~8 for FIR1 (Project ID: 2021.1.01706.S; PI: Y. Cheng; see \citealt{Cheng25} for details), while the other regions were observed in Cycle~11 (Project ID: 2024.1.01580.S; PI: Y. Cheng). The six regions cover areas ranging from approximately 5 to 20~$\mathrm{arcmin}^2$. The observations used the 12~m array in the C43-3 configuration for FIR1 and in the C43-4/C43-1 configurations for the other regions, together with the 7~m array to recover scales up to $\sim29\arcsec$. Total-power observations were also obtained for the recovery of larger-scale line emission.

The spectral setup was the same for all regions and is described in \citet{Cheng25}. {In this paper, we use the continuum spectral window with a bandwidth of 1.875~GHz centered at 231.00~GHz.} The raw data were calibrated with the ALMA data-reduction pipeline using \texttt{CASA} 6.2.1. The continuum visibility data were constructed from line-free channels. We imaged the 12~m and 7~m array data using the \texttt{tclean} task in \texttt{CASA} using Briggs weighting with a robust parameter of 0.5. The resulting angular resolution is approximately 0.6--0.8\arcsec{} across different fields, and the rms sensitivities range from 0.11 to 0.13~\mjypbm (see \autoref{table:sum}).

In addition, we also derive large scale dust temperature and column density maps from archival \textit{Herschel} 70--500~$\mu$m images by fitting a single temperature modified blackbody model, as described in detail in \autoref{sec:sed}.

\begin{figure*}[ht!]
\epsscale{1.1}\plotone{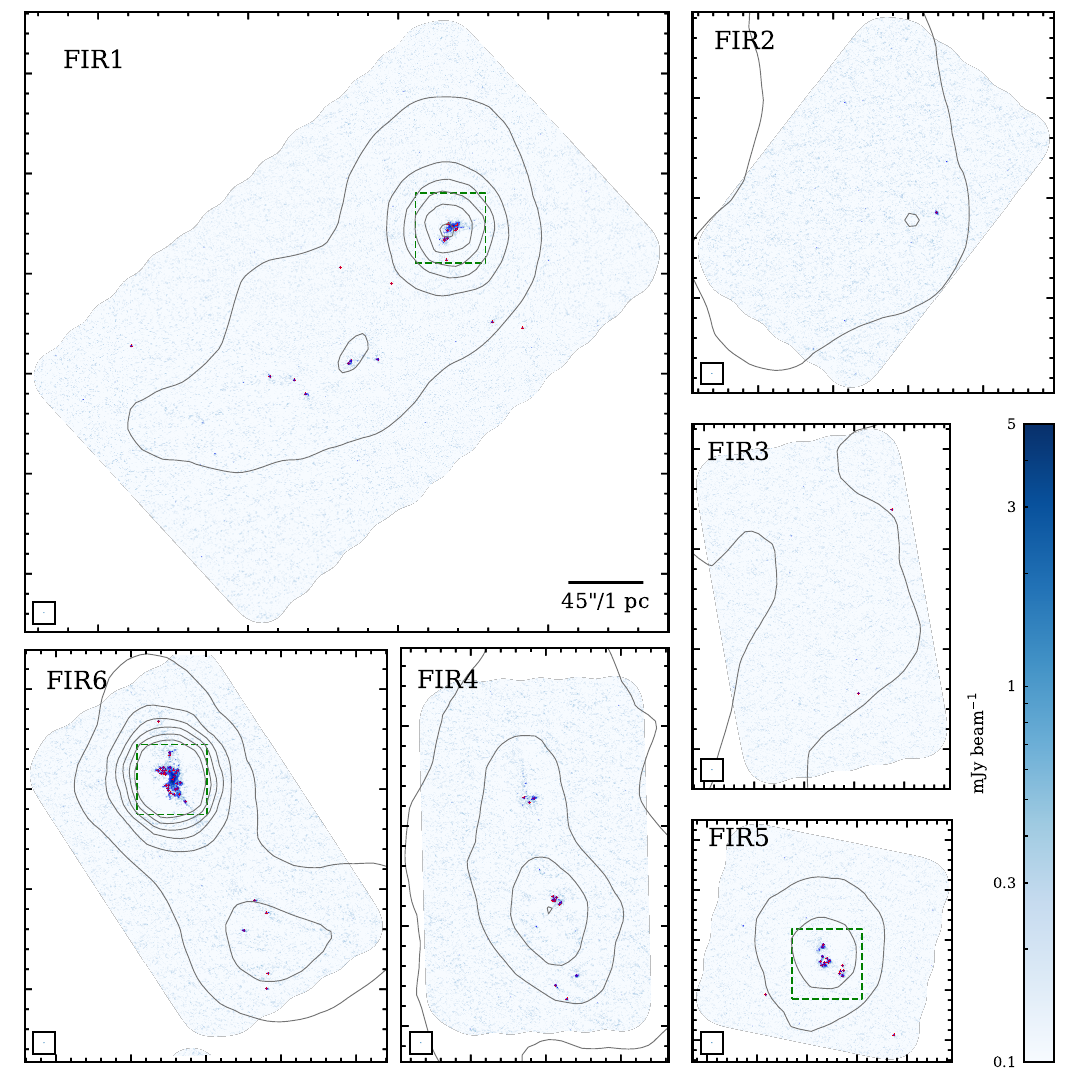}
\caption{ALMA 1.3~mm continuum images of the six targeted regions in Sh2-284. The blue color scale and blue contours show the ALMA continuum emission, with contour levels of $(4,8,16,32,64)\times\sigma$, where $\sigma$ is the rms sensitivity of each field listed in \autoref{table:sum}. Gray contours show the {\it Herschel} 500~$\mu$m emission at $(30,60,90,120,150,180)$~MJy~sr$^{-1}$, tracing the larger-scale cloud structure. The green dashed boxes indicate the regions shown in \autoref{fig:cont_zoom}. All panels are displayed with the same physical scale; the scale bar and color bar are shared by all panels.
}\label{fig:cont}
\end{figure*} 

\begin{deluxetable*}{cccccccccc}
\tablecaption{Properties of the Observed Regions in Sh2-284 \label{table:sum}}
\tablehead{
\colhead{Region} &
\colhead{R.A.} &
\colhead{Decl.} &
\colhead{Size} &
\colhead{$N_{\rm H_2}$} &
\colhead{$T_{\rm dust}$} &
\colhead{Beam} &
\colhead{Sensitivity} &
\colhead{$N_{\rm core}$} &
\colhead{$M_{\rm core,med}$} \\
\colhead{} &
\colhead{(J2000)} &
\colhead{(J2000)} &
\colhead{(\arcmin{} $\times$ \arcmin{})} &
\colhead{($10^{21}$ cm$^{-2}$)} &
\colhead{(K)} & 
\colhead{(\arcsec{} $\times$ \arcsec{})} &
\colhead{(mJy beam$^{-1}$)} & 
\colhead{} &
\colhead{$M_\odot$} 
}
\startdata
\hline
FIR1 & 06:46:20.19 & +00:05:33.4 & 3.5 $\times$ 5.7 & 6.1 (22.6) & 16.4 (22.0) & 0.64 $\times$ 0.54 & 0.11 & 22/21 & 2.1/2.9 \\
FIR2 & 06:46:01.66 & +00:14:01.6 & 2.2 $\times$ 3.4 & 6.9 (12.9) & 17.8 (18.8) & 0.69 $\times$ 0.57 & 0.13 & 1/1 & 5.7/6.5 \\
FIR3 & 06:45:59.37 & +00:18:25.4 & 2.2 $\times$ 3.3 & 6.2 (8.4) & 18.5 (19.8) & 0.78 $\times$ 0.55 & 0.11 & 2/2 & 1.2/1.9 \\
FIR4 & 06:45:15.47 & +00:25:43.2 & 2.2 $\times$ 3.5 & 8.6 (19.1) & 18.5 (20.1) & 0.78 $\times$ 0.60 & 0.12 & 13/9 & 1.2/2.5 \\
FIR5 & 06:44:49.44 & +00:20:30.1 & 2.2 $\times$ 2.1 & 4.6 (11.1) & 19.5 (21.8) & 0.76 $\times$ 0.60 & 0.11 & 17/12 & 0.9/2.1 \\
FIR6 & 06:47:12.29 & +00:25:27.2 & 2.3 $\times$ 3.6 & 6.5 (35.9) & 16.8 (26.8) & 0.78 $\times$ 0.60 & 0.13 & 36/23 & 1.7/3.4 \\
\hline
\enddata
\tablecomments{The $N_{\rm H_2}$ and $T_{\rm dust}$ values are derived from the \textit{Herschel} SED fits within each region footprint and are reported as the median, with the peak value within the region given in parentheses. $N_{\rm core}$ gives the number of cores in the full catalog and in the robust catalog, respectively. $M_{\rm core,med}$ gives the corresponding median core masses for the two catalogs, estimated from the 1.3~mm fluxes as described in \autoref{sec:core}.}
\end{deluxetable*}

\begin{figure*}[ht!]
\epsscale{1.1}\plotone{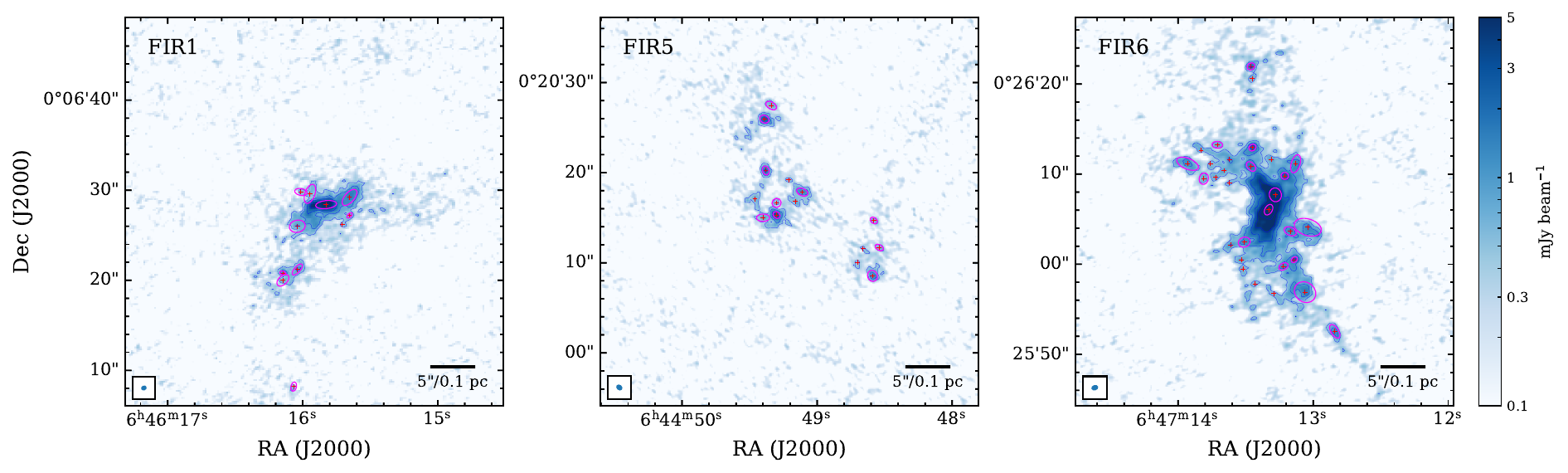}
\caption{Zoom-in views of the ALMA 1.3~mm continuum emission toward the clustered regions in FIR1, FIR5, and FIR6. The blue color scale and blue contours show the continuum emission, with contour levels the same as in \autoref{fig:cont}. Red crosses mark the cores identified in the full dendrogram catalog, while magenta ellipses show the Gaussian components for cores retained in the robust dendrogram+Gaussian catalog. The synthesized beam and scale bar are shown in the bottom-left and bottom-right corners of each panel, respectively.
}\label{fig:cont_zoom}
\end{figure*}

\section{Results}\label{sec:results}

\subsection{1.3~mm Continuum}

\autoref{fig:cont} presents the 1.3~mm continuum images of the six regions, overlaid with \textit{Herschel} 500~$\mu$m contours tracing the large scale emission. The regions show very different levels of fragmentation on $\sim$3000~au scales. Clustered groups of cores are seen in the northwestern part of FIR1, the northeastern part of FIR6, and FIR5, with $\sim$10--20 cores within $\sim$1~pc; these regions are generally associated with strong \textit{Herschel} 500~$\mu$m emission and are shown in greater detail in \autoref{fig:cont_zoom}. In other regions, the cores are more sparsely distributed. In particular, FIR2 and FIR3, which have the weakest 500~$\mu$m emission in the sample, show almost no detected dense structures on $\sim$3000~au scales. Overall, the number of dense cores appears to be closely linked to the parsec-scale cloud conditions.

\subsection{Identification and characterization of cores}\label{sec:core}

We use the {\it astrodendro} algorithm \citep{Rosolowsky08} to search for dense cores. This algorithm is designed for identifying hierarchical structures in continuum or line data and we define the leaves, i.e., the base element in the dendrogram hierarchy without further substructures, as cores. We adopt a base flux density threshold of 4$\sigma$, a minimum significance of 1$\sigma$, and a minimum area of half the synthesized beam. The same parameter choices have been used in previous CMF studies \citep[][]{Cheng18,Liu18,ONeill21,Kinman25}. Cores are identified on the maps before primary beam correction, while their flux densities are measured on the corrected maps by summing over the pixels assigned to each leaf.

{\it Astrodendro} detects 91 cores with the adopted criteria. The identified cores are marked in \autoref{fig:cont} and \autoref{fig:cont_zoom}. The number of cores varies among the six regions, ranging from 1 in FIR2 to 36 in FIR6 (see \autoref{table:sum}). Visual inspection shows that some faint leaves returned by \texttt{astrodendro} have low contrast and may partly reflect local background fluctuations rather than compact cores. This is expected because \texttt{astrodendro} is a pixel-assignment-based algorithm and does not explicitly subtract large scale or background emission; as a result, small 1--2$\sigma$ fluctuations on top of extended emission above the $4\sigma$ threshold can satisfy the adopted criteria. We therefore retain this initial list as a full catalog of potential cores, and further define a robust catalog using two-dimensional Gaussian fitting. For each leaf, we fit a 2D Gaussian together with a constant background level as a free parameter, and include only sources with fitted Gaussian amplitudes exceeding $5\sigma$ in the robust catalog. This selection yields 68 robust cores, with those in the clustered regions marked in \autoref{fig:cont_zoom}.

We estimate core gas masses assuming optically thin, isothermal dust emission:
\begin{equation}
M_{\rm gas} = R_{\rm gd}\frac{d^2F_\nu}{\kappa_\nu B_\nu(T_{d})},
\end{equation}
where $d$ is the distance of 4.5~kpc, $R_{\rm gd}$ is the gas-to-dust mass ratio, $F_\nu$ is the observed flux density, $B_\nu(T_d)$ is the Planck function at dust temperature $T_d$, and $\kappa_\nu$ is the dust opacity at the observed frequency. We assume a uniform dust temperature of 20~K and adopt $\kappa_{\rm 1.3mm}=0.899~{\rm cm^2~g^{-1}}$, corresponding to the moderately coagulated thin-ice-mantle dust model of \citet{Ossenkopf94}. We adopt $R_{\rm gd}=282$, obtained by scaling the local Galactic gas-to-dust mass ratio of 141 \citep{Draine11} by the inverse of the assumed half-solar metallicity. The resulting core masses range from $0.37$ to $90~\msun$ for the full catalog, based on the dendrogram-assigned flux densities, and from $1.0$ to $57~\msun$ for the robust catalog, based on the Gaussian-fit fluxes. A full catalog of the core properties is listed in \autoref{sec:core_cat}.

\subsection{Core mass function (CMF)}\label{sec:cmf}

\begin{figure*}[ht!]
\epsscale{1.1}\plotone{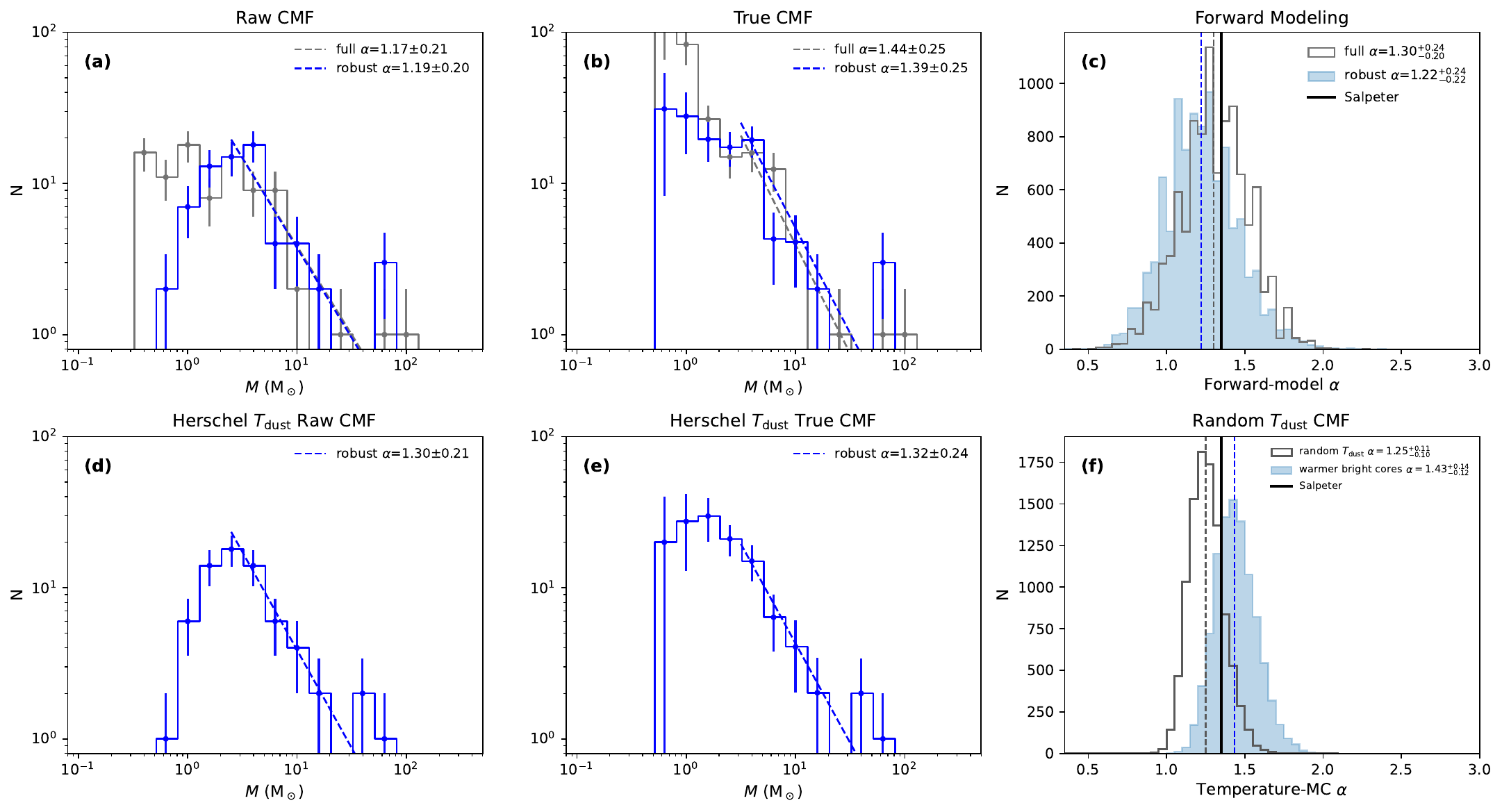}
\caption{Core mass functions of Sh2-284. \textit{(a)} Raw CMFs derived from the full dendrogram catalog and the robust dendrogram+Gaussian catalog, assuming a uniform dust temperature of 20~K. Dashed lines show the maximum-likelihood estimation (MLE) power-law fits above $M>2.5~\msun$. \textit{(b)} Recovery-corrected true CMFs; for the full catalog, both flux and number corrections are applied, while for the robust catalog only the number correction is applied. Dashed lines show the binned-MLE power-law fits above $M>2.5~\msun$. {With
the adopted logarithmic binning, the first included bin has a lower
edge of $3.162~M_\odot$.} \textit{(c)} Forward-modeling distributions of the intrinsic CMF slope for the two catalogs; the vertical solid line marks the Salpeter value, $\alpha=1.35$. {\textit{(d)} Raw CMF for the robust catalog after assigning each core the local
Herschel-derived dust temperature. The dashed line shows the
MLE power-law fit for $M>2.5~\msun$. 
\textit{(e)} Number-corrected CMF for the same temperature assignment. The dashed line shows the binned-MLE power-law fits for $M>2.5~\msun$. }
\textit{(f)} Monte Carlo temperature test in which core temperatures are
randomly varied around the local \textit{Herschel} values, with an additional
warmer-bright-core case to test unresolved internal heating (see \autoref{sec:robustness} for details). The vertical line
marks the Salpeter slope.
}\label{fig:cmf}
\end{figure*}

\begin{deluxetable*}{cccccc}
\tabletypesize{\scriptsize}
\tablecaption{CMF slope measurements for different lower fitting
limits and core temperature treatments
\label{table:cmf_fits}}
\tablehead{
\colhead{Catalog} &
\colhead{Core temperature} &
\colhead{Fitting method} &
\colhead{$M_{\rm min}$ ($M_\odot$)} &
\colhead{${N_{\rm fit}}^{b}$} &
\colhead{$\alpha$}
}
\startdata
Full catalog, raw & Fixed 20 K & MLE & 2.0, fixed & 38 & $1.15\pm0.19$ \\
& & & 2.5, fixed & 30 & $1.17\pm0.21$ \\
& & & 3.0, fixed & 24 & $1.18\pm0.24$ \\
& & & 2.11, KS-selected$^{a}$ & 38 & $1.22\pm0.20$ \\
Robust catalog, raw & Fixed 20 K & MLE & 2.0, fixed & 46 & $1.13\pm0.17$ \\
& & & 2.5, fixed & 37 & $1.19\pm0.20$ \\
& & & 3.0, fixed & 33 & $1.34\pm0.23$ \\
& & & 2.36, KS-selected$^{a}$ & 43 & $1.29\pm0.20$ \\
Full catalog, flux and number corrected & Fixed 20 K & Binned MLE & 1.995, fixed & 47 [50.44] & $1.14\pm0.16$ \\
& & & 3.162, fixed & 34 [35.47] & $1.44\pm0.25$ \\
& & & 3.162, KS-selected$^{a}$ & 34 [35.47] & $1.44\pm0.25$ \\
Robust catalog, number corrected & Fixed 20 K & Binned MLE & 1.995, fixed & 46 [50.13] & $1.18\pm0.17$ \\
& & & 3.162, fixed & 31 [32.78] & $1.39\pm0.25$ \\
& & & 1.995, KS-selected$^{a}$ & 46 [50.13] & $1.18\pm0.17$ \\
Robust catalog, raw & \textit{Herschel} $T_{\rm dust}$ & MLE & 2.0, fixed & 47 & $1.16\pm0.17$ \\
& & & 2.5, fixed & 40 & $1.30\pm0.21$ \\
& & & 3.0, fixed & 30 & $1.21\pm0.22$ \\
& & & 2.44, KS-selected$^{a}$ & 42 & $1.32\pm0.20$ \\
Robust catalog, number corrected & \textit{Herschel} $T_{\rm dust}$ & Binned MLE & 1.995, fixed & 47 [51.51] & $1.24\pm0.17$ \\
& & & 3.162, fixed & 29 [30.47] & $1.32\pm0.24$ \\
& & & 1.995, KS-selected$^{a}$ & 47 [51.51] & $1.24\pm0.17$ \\
\enddata
\tablenotetext{a}{The KS-selected lower limit is chosen by minimizing the KS distance between
the observed cumulative mass distribution and the fitted power-law
model after restricting candidate lower limits for mass bins with completeness above 80\%. For the binned MLE, the data are represented by counts in fixed 0.2-dex logarithmic mass bins. Consequently, the fitted lower limit is necessarily the lower edge of the first included bin (e.g., $1.995$, $3.162$, or $5.012~M_\odot$), rather than an arbitrary intra-bin mass value.}
\tablenotetext{b}{Bracketed values of $N_{\rm fit}$ are recovery-weighted effective counts.}
\end{deluxetable*}

We construct the CMFs for both the full and robust catalogs in \autoref{fig:cmf}. {We consider two corrections that may influence the measured CMFs.} First, the detection completeness (or number recovery fraction, $f_{\rm num}$) depends on core fluxes, and therefore directly affects the number distribution in each mass bin. {Second, the dendrogram flux measurement can be affected by two competing effects. Because the flux is summed only over pixels assigned
to each leaf, low intensity source emission outside the isophotal leaf boundary is not included. On the other hand, because no background emission is subtracted, extended large scale emission within the leaf aperture may increase the measured flux. We thus define $f_{\rm flux}$ as an empirical flux correction factor that accounts for the net bias introduced by the dendrogram extraction procedure.}

To derive the correction factors, we perform artificial core injection experiments, as described in detail in \autoref{sec:core_insert}. Briefly, we generate artificial cores with a range of flux densities or masses, assigning realistic sizes and shapes empirically determined from the observed core population. The artificial core positions are drawn from a two-dimensional kernel density estimate (KDE) constructed from the observed core positions in each field, so that the injections follow the spatial distribution of observed cores. For each flux density, we inject five artificial cores per region in each realization and repeat the experiment 1000 times. We then run the same core identification and flux measurement procedures used for the real data to determine both the number-recovery fraction, $f_{\rm num}$, and the flux-recovery fraction, $f_{\rm flux}$. We treat the full dendrogram catalog and the robust dendrogram+Gaussian catalog as two separate catalog-construction methods. The latter applies an additional local contrast criterion and measures fluxes from Gaussian fits rather than dendrogram-assigned pixels. The resulting correction factors for the two methods are discussed in \autoref{sec:core_insert}. Overall, the $f_{\rm flux}$ curve for the dendrogram+Gaussian method remains close to unity, indicating more stable flux recovery. We therefore apply the flux correction only to the pure dendrogram catalog, while the robust catalog is corrected only for number recovery.

As an initial characterization, we fit the high-mass end ($M>2.5~\msun$) of the raw CMF with a power-law function,
\begin{equation}
\frac{dN}{d\log M} \propto M^{-\alpha},
\label{equ}
\end{equation}
using the maximum-likelihood estimation (MLE) method of \citet{Clauset09}, as implemented in the \texttt{plfit} package. This threshold of 2.5~\msun{} corresponds to a typical injected peak significance of $\sim$10$\sigma$ for the empirical core-size distribution adopted in the insertion experiments, and to $\gtrsim$80\% detection completeness (see \autoref{sec:core_insert}), providing a good compromise between rejecting false detections and retaining sufficient statistics for the fit. This gives slopes of $\alpha=1.17\pm0.21$ and $\alpha=1.19\pm0.20$ for the two catalogs, shown in \autoref{fig:cmf}a and \autoref{table:cmf_fits}. 

{We also test the sensitivity of the inferred slope to different
fitting limits. For the raw CMFs, we repeat the MLE fits using fixed
values of $M_{\rm min}=2.0$, 2.5, and $3.0~M_\odot$. 
We also estimate $M_{\rm min}$ by minimizing the KS distance between
the observed cumulative mass distribution and the fitted power-law
model. Candidate lower limits are considered only within the mass
range where the number completeness exceeds 80\%.
The KS-selected lower limits are $2.11$ and $2.36~M_\odot$ for the
full and robust catalogs, respectively. As summarized in
\autoref{table:cmf_fits}, the fitted slopes remain consistent
within their uncertainties for all tested lower limits. We retain $M_{\rm min}=2.5~M_\odot$ as a common fiducial threshold in the following analysis, while the results obtained with alternative fixed and KS-selected lower limits are summarized in \autoref{table:cmf_fits}.}

We then correct the measured masses of the catalog using the corresponding $f_{\rm flux}$ values (for the full catalog), and then correct the number counts in each mass bin using $f_{\rm num}$ (for both catalogs) to get the ``true'' CMF. Because the number-corrected CMF is defined only through binned counts, the standard unbinned MLE cannot be applied directly. We therefore use the binned MLE of \citet{Virkar14} to fit the number-corrected CMF. {For consistency with the fiducial raw-CMF threshold,
we include only bins lying entirely above $2.5~M_\odot$. With the
adopted 0.2-dex logarithmic binning, the first such bin has a lower
edge of $3.162~M_\odot$.} The resulting slopes are
$\alpha=1.44\pm0.25$ and $\alpha=1.39\pm0.25$ for the full and robust
catalogs, respectively. {We also repeat the binned fits using alternative lower bin edges and select the preferred $M_{\rm min}$ through KS minimization. Since the data are binned, $M_{\rm min}$ is restricted to the adopted bin edges. As summarized in \autoref{table:cmf_fits}, the resulting slopes remain consistent within their uncertainties, although the fits beginning at $3.162~M_\odot$ yield slightly steeper values.}

Although the binned MLE provides a likelihood-based treatment of binned power-law distributions, it cannot recover the information lost during the binning process, where individual core masses are replaced by counts within finite mass intervals. We therefore further adopt a forward-modeling approach to constrain the intrinsic CMF slope. The idea is to ask what the observed cumulative mass distribution would look like if the underlying CMF were a power law, after accounting for the fact that cores of different masses are recovered with different probabilities and, for dendrogram-only method, with different flux recovery fractions. For each trial slope $\alpha$ from 0 to 2 in step of 0.01, where ${\rm d}N/{\rm d}\log M\propto M^{-\alpha}$, we represent the intrinsic CMF on a fine logarithmic mass grid over $2.5\leq M/M_\odot\leq100$. Each grid point represents a small logarithmic mass interval, and is assigned a relative expected number of cores proportional to $M^{-\alpha}$. This model number at each mass is then multiplied by the number recovery fraction, $f_{\rm num}(M)$, because only this fraction of cores with that true mass would be detected in the artificial-core experiments. Thus, the weights on the mass grid represent the expected observed number distribution for a given input slope. For dendrogram-only method, we also account for flux loss by mapping each true mass on the grid to the corresponding measured mass, $M_{\rm obs}=M\cdot f_{\rm flux}(M)$; for the dendrogram+Gaussian method, the masses are left unchanged and only the number recovery is applied. The resulting model points, now described by an observed mass and an expected observed number weight, are sorted by observed mass and cumulatively summed to form the model cumulative distribution function. This procedure is applied separately to each field using its own recovery curves, and the field-specific cumulative distributions are then combined, weighted by the number of observed cores in the fitted mass range.

We then compare this model cumulative distribution with the cumulative distribution of the observed core masses using a Kolmogorov--Smirnov statistic. The KS statistic measures the maximum vertical separation between the observed cumulative mass distribution and the model prediction for a given input slope. The best-fit slope is the trial value that minimizes this separation. To estimate the uncertainty, we bootstrap the observed cores within each field, repeat the KS comparison for each bootstrap realization, and record the best-fit slope. The resulting distribution of best-fit slopes is shown in \autoref{fig:cmf}c. We report the slope as $1.30^{+0.24}_{-0.20}$ and $1.22^{+0.24}_{-0.22}$ for the two catalogs, using the median and 68\% central interval.


\subsection{Robustness of the CMF} \label{sec:robustness}

The derived CMF depends on the assumptions made when converting 1.3~mm flux density into core mass. For the gas-to-dust ratio, we adopt $R_{\rm gd}=282$, obtained by scaling the Galactic reference value of 141 \citep{Draine11} by the inverse of the assumed half-solar metallicity. Since the metallicity of Sh2-284 has been estimated to be $\sim1/3$--$1/2 Z_\odot$ \citep{Negueruela15}, adopting the lower end of this range would increase all core masses by a factor of 1.5. This would shift the CMF horizontally but does not affect the measured high-mass slope.

Overall, Sh2-284 appears to be dominated by relatively low- to intermediate-mass star formation. Even within the targeted fields, the {\it Herschel}-derived column densities typically lie in the range $N_{\rm H_2}\sim10^{21}$--$10^{22}{\rm cm^{-2}}$ (see \autoref{sec:sed}), comparable to nearby low mass clouds such as Taurus \citep{Andre10}. Only one or two localized structures, most notably the dense clump in FIR6, reach column densities exceeding $10^{23} {\rm cm^{-2}}$, where conditions may be more favorable for forming higher-mass stars; FIR1 is another region where high-mass stars may be forming \citep{Cheng25}. Therefore, we do not expect free-free emission to affect the CMF at the population level. 

We also test the effect of temperature variations using the {\it Herschel}-derived dust temperature map (\autoref{fig:cmf}d,e). For the robust catalog, we assign each core the dust temperature at its position and re-derive the CMF. Across the six regions studied here, the dust temperature of cores ranges from 15.0~K to {25.2~K} with a median of {20.8~K}. Because the source detection and completeness depend on flux density rather than mass, the number recovery correction is applied as a function of the observed core flux. 
{Adopting a lower mass threshold of 
$M_{\min}=2.5\,M_\odot$, we apply the unbinned MLE to the 
individual core masses above this limit for the raw 
Herschel-temperature CMF. For the number-corrected CMF that is 
defined using binned counts, we apply the binned MLE beginning at 
the first logarithmic bin lying above $2.5\,M_\odot$, 
corresponding to a lower bin edge of $3.162\,M_\odot$. The resulting 
raw and number-corrected slopes are $\alpha=1.30\pm0.21$ and 
$\alpha=1.32\pm0.24$, respectively. Results obtained using alternative 
fixed and KS-selected lower mass limits are also listed in 
\autoref{table:cmf_fits}.} These slopes are consistent with the 
fiducial result derived assuming a uniform temperature.

Since the {\it Herschel}-derived temperatures are measured at much coarser
resolution than the ALMA cores and may not capture local heating by embedded
protostars, we further test the impact of uncertain core temperatures with a
Monte Carlo resampling experiment. For each realization, we assign each robust
core a random dust temperature drawn uniformly from
$T_{\rm Herschel}-5~{\rm K}$ to $T_{\rm Herschel}+15~{\rm K}$, convert the
observed 1.3 mm fluxes to core masses, apply the same flux-dependent number
recovery correction, and repeat the binned-MLE fit starting
from the first mass bin with mean completeness above 80\%. This yields a distribution of $\alpha$, with $\alpha = {1.25^{+0.11}_{-0.10}}$ determined from the median and the central 68\% interval. As a more extreme test for unresolved internal heating, we repeat the same experiment but assign the brightest cores, defined as those with $M_{T_{\rm Herschel}}>10~M_\odot$, warmer temperatures
drawn uniformly from 30 to 70 K. This warmer-bright-core case gives
$\alpha={1.43^{+0.14}_{-0.12}}$. Both distributions are shown in
\autoref{fig:cmf}f. Thus, plausible core-to-core temperature variations shift
the inferred slope slightly but remain consistent with a
Salpeter-like high mass CMF.


\section{Discussion}

The measured high-mass CMF slope in Sh2-284, $\alpha=1.22^{+0.24}_{-0.22}$, is statistically consistent with the Salpeter slope of $\alpha=1.35$ \citep{Salpeter55}. A comparison with CMFs measured using a uniform analysis framework across different Galactic environments is presented in \autoref{sec:cmf_comp}. In the context of a possible CMF--IMF connection, it is useful to compare this result with IMF measurements in other low metallicity environments. Previous near-infrared studies of low metallicity clusters in the outer Galaxy reached stellar masses of order $\sim0.1$--$0.2~M_\odot$ and favored canonical Galactic IMFs based on fitting the $K$-band luminosity functions \citep{Yasui08,Yasui16a,Yasui16b,Yasui21}. In the Magellanic Clouds, several studies of young clusters and associations have also found high- or intermediate-mass stellar mass functions consistent with canonical forms for $M\gtrsim1~M_\odot$ \citep[e.g.,][]{Schmalzl08,Liu09,DaRio09,Andersen09,DaRio12,Cohen26}. Within Sh2-284 itself, optical/near-infrared studies of Dolidze~25 have found stellar mass functions broadly consistent with a Salpeter/Kroupa form over the low- to intermediate-mass pre-main sequence (PMS) regime: \citet{Delgado10} reported a Salpeter-like mass function over $\sim1.3$--$3.5~M_\odot$, while \citet{Guarcello21}, using X-ray and infrared-selected members, found no significant deviation from a Salpeter slope over $\sim0.8$--$2~M_\odot$. Therefore, the CMF slope found here is consistent with the available IMF constraints in Sh2-284. This resembles the empirical CMF--IMF correspondence observed in nearby Galactic clouds \citep[e.g.,][]{Motte98,Alves07,Konyves15}, suggesting that the shape of IMF at the high mass end may already be imprinted in the dense core population in a metal-poor environment in the outer Galaxy.

This picture is not universal, however. A shallower slope or more top-heavy IMF has been reported in some moderately metal-poor environments, including 30~Doradus, where the massive-star IMF over $15$--$200~M_\odot$ is significantly shallower than Salpeter \citep{Schneider18}; similar tendencies have also been suggested in Sh2-209 and NGC~796 \citep{Yasui23,Kalari18}. These results imply that the high-mass IMF is likely regulated by other environmental factors in addition to the metallicity. Interestingly, recent ALMA observations of 30Dor-10 find a Salpeter-like CMF at $\sim2000$~au resolution over core masses of approximately 10--100~\msun, despite the top-heavy massive-star IMF inferred for the broader 30~Dor region \citep{Traficante26}. This suggests that the observed CMF may not map directly onto the final IMF, with later core evolution and stellar mass assembly playing an important role. As discussed in \autoref{sec:robustness}, Sh2-284 samples a regime of relatively quiescent low- to intermediate-mass star formation in the low surface density regions, rather than the high-pressure, massive cluster environment represented by 30~Dor. Its Salpeter-like CMF therefore provides a useful low metallicity comparison case in which the CMF--IMF resemblance appears closer to that seen in solar neighbor clouds. In more extreme environments, the core population may undergo more substantial evolution after the observed CMF stage, leading to a more complex mapping between the CMF and the final IMF \citep{Traficante26}.

While both the CMF and IMF in Sh2-284 are Salpeter-like at the high-mass end, the JWST/NIRCam census of S284-EC1 (FIR1 region in this work) suggests a lower IMF turnover mass of $m_{\rm c}=0.16\pm0.02 M_\odot$ than typically measured in local young clusters \citep{Andersen25}. Similar low characteristic or turnover masses have also been reported in other metal-poor regions in the outer Galaxy and the Magellanic Clouds \citep{Yasui24,Cohen26}. These results may hint that metallicity-dependent effects, if present, may be more readily identified through the mass scale of the IMF than through its high-mass slope. One likely interpretation is that the high-mass slope reflects a broad combination of fragmentation, accretion, and dynamical evolution, whereas the turnover mass is tied to a characteristic fragmentation scale set by thermal balance, dust opacity, and radiative heating, where metallicity may play a larger role. Our ALMA data are not sensitive enough to probe the expected CMF turnover regime, so it is unclear whether the low stellar characteristic mass is already imprinted in the core population.

\section{Summary}
We have measured the CMF across six FIR-bright regions in Sh2-284 with ALMA 1.3~mm observations at $\sim3000$~au resolution. The core population is highly non-uniform, with clustered groups in FIR1, FIR5, and FIR6 but little compact emission in FIR2 and FIR3. Using both a full dendrogram catalog (91 cores) and a robust dendrogram+Gaussian catalog (68 cores), and correcting for number completeness and flux recovery, we find Salpeter-like high-mass CMF slopes. Forward modeling gives $\alpha=1.30^{+0.24}_{-0.20}$ and $\alpha=1.22^{+0.24}_{-0.22}$ for the two catalogs, respectively, and tests with \textit{Herschel}-based and Monte Carlo dust temperatures yield consistent slopes. Together with the available IMF constraints in Sh2-284, this suggests that the high-mass CMF--IMF resemblance seen in solar-neighborhood clouds can also be present in a low metallicity outer Galaxy environment. The next key step is to push the CMF measurement to lower masses, where the connection to the possibly low stellar characteristic mass can be tested directly.

\acknowledgments
We thank S. H. Jiao for helpful discussions regarding the Herschel SED fitting. This paper makes use of the following ALMA data: ADS/JAO.ALMA\#2021.1.01706.S and 2024.1.01580.S. ALMA is a partnership of ESO (representing its member states), NSF (USA) and NINS (Japan), together with NRC (Canada), MOST and ASIAA (Taiwan), and KASI (Republic of Korea), in cooperation with the Republic of Chile. The Joint ALMA Observatory is operated by ESO, AUI/NRAO and NAOJ. The National Radio Astronomy Observatory is a facility of the National Science Foundation operated under cooperative agreement by Associated Universities, Inc. Y.C. was partially supported by Grant-in-Aid for Scientific Research (KAKENHI number 24K17103 and 26K00748) of the JSPS. K.T. was partially supported by Grant-in-Aid for Scientific Research (KAKENHI number 25K07365) of the JSPS. K.T. acknowledge support from the NAOJ ALMA Scientific Research Grant Code 2025-29B. Data analysis was in part carried out on the Multi-wavelength Data Analysis System operated by the Astronomy Data Center (ADC), National Astronomical Observatory of Japan. R.F. acknowledges support from the grant PID2023-146295NB-I00, and from the Severo Ochoa grant CEX2021-001131-S funded by MCIN/AEI/10.13039/501100011033 and by ``European Union NextGenerationEU/PRTR''.

\software{CASA \citep{Mcmullin07,casa22}, APLpy \citep{Aplpy12}, Astropy \citep{Astropy13,astropy18,astropy22}}

\clearpage
\appendix
\counterwithin{figure}{section}
\counterwithin{table}{section}

\section{Herschel SED fit}\label{sec:sed}

For the spectral energy distribution (SED) fits, we assume that the far-infrared emission in each pixel is described by a single-component modified black-body. Before performing the fitting, all Herschel images at 70, 160, 250, 350, and 500$\mu$m were smoothed to the common angular resolution of the SPIRE 500$\mu$m beam, $\sim$37\arcsec, and re-gridded to the 500$\mu$m pixel grid. 

The flux density is given by
\begin{equation}
S_\nu = \Omega_mB_\nu(T_\mathrm{d})(1-e^{-\tau_\nu}),
\end{equation}
and the column density is
\begin{equation}
N_\mathrm{H_2} = \frac{\tau_\nu M_g}{\kappa_\nu \mu m_\mathrm{H}M_d},
\end{equation}
where $B(T_d)$ is the Planck function at $T_d$, $\mu$ is the mean molecular weight per H$_2$ molecule of 2.8, and $\kappa$=$\kappa_{230}(\nu/230 {\mathrm GHz})^\beta$, $\Omega_{m}$ is the solid angle. We adopt $\kappa_{230}$ = 0.899 $\rm cm^2g^{-1}$ as in \citet{Ossenkopf94} for the moderately coagulated thin ice mantle model. We fix $\beta=1.8$. To account for the sub-solar metallicity of Sh2-284, we adopt a gas-to-dust ratio $M_g/M_d=282$, corresponding to $141/Z$ with $Z=0.5$.

\begin{figure*}[ht!]
\epsscale{1.1}\plotone{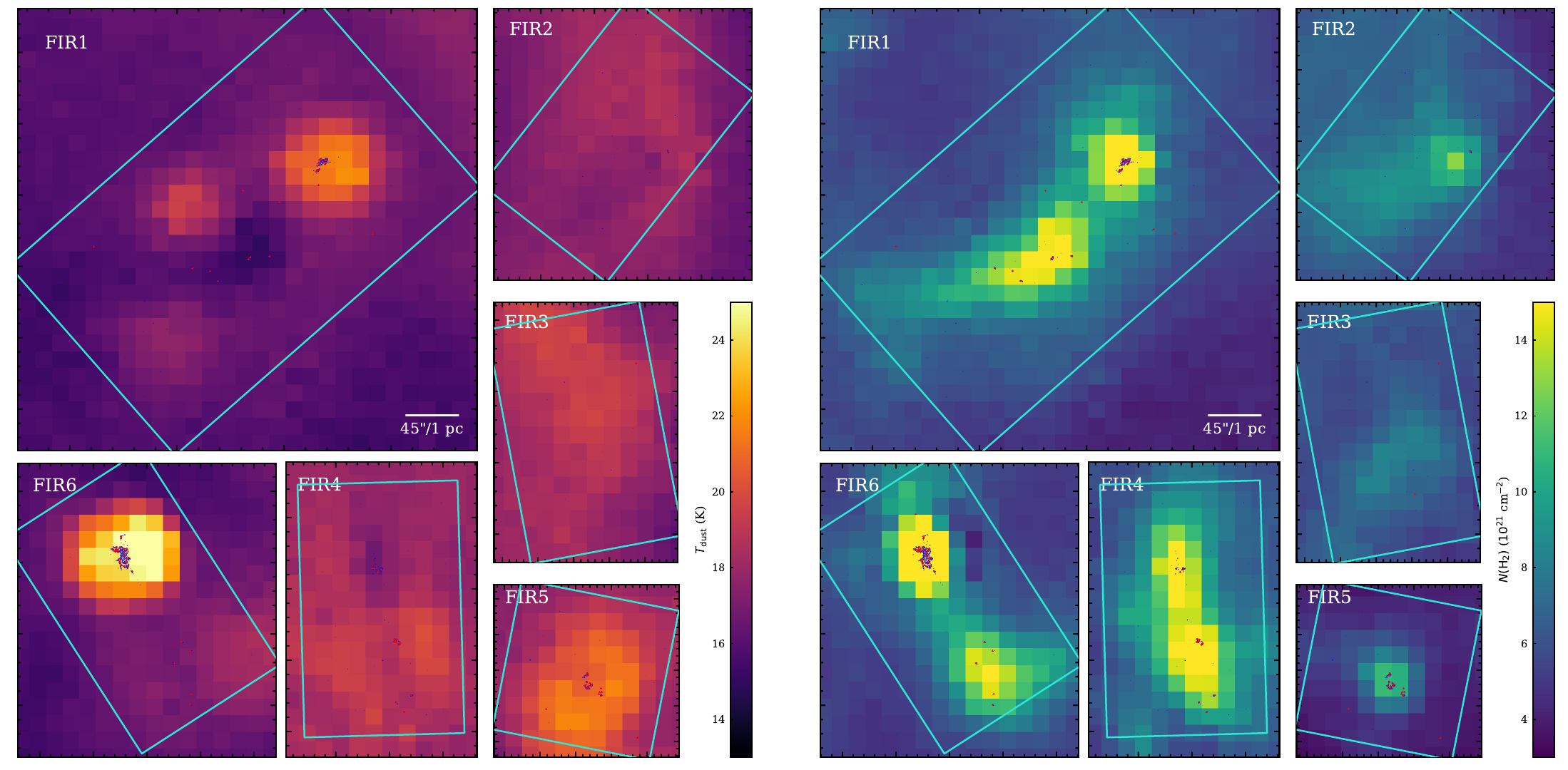}
\caption{Dust temperature $T_{\rm dust}$ and column density $N_{\rm H_2}$ map from the SED fit for the six regions. The blue contours and red crosses show the ALMA cores.
}\label{fig:sed}
\end{figure*}

We performed pixel-by-pixel least-squares fits to the 160/250/350/500$\mu$m SEDs. The fits were weighted by the inverse variance of the Herschel uncertainty maps. The error maps were smoothed and re-gridded in the same way as the science images, converted to Jy pixel$^{-1}$ on the common grid, and a median positive-error floor was applied in each band. {We excluded the 70~$\mu$m data from the fitting, since the 70~$\mu$m emission can contain excess contribution from warmer dust heated by embedded protostars that is not well described by a single temperature component \citep{Elia17}. Our tests also show that excluding 70~$\mu$m yields improved fit in the target regions.} Pixels were fitted when at least four bands had finite positive fluxes and uncertainties. The resulting column density and dust temperature maps have an angular resolution of $\sim$37\arcsec, and are shown in \autoref{fig:sed}.

\section{Core catalog}\label{sec:core_cat}

In \autoref{table:core_cat} we list the properties of the identified cores.

\begin{deluxetable}{cccccccccccccc}
\tabletypesize{\scriptsize}
\setlength{\tabcolsep}{2pt}
\tablecaption{Properties of the Identified Dense Cores in Sh2-284 \label{table:core_properties}}
\tablehead{
\colhead{Region} &
\colhead{Core} &
\colhead{R.A.} &
\colhead{Decl.} &
\colhead{$S_{\nu,{\rm dendro}}$} &
\colhead{$M_{\rm dendro,20K}$} &
\colhead{Area} &
\colhead{$S_{\nu,{\rm gauss}}$} &
\colhead{$M_{\rm gauss,20K}$} &
\colhead{$I_{\nu,{\rm gaussian,peak}}$} &
\colhead{FWHM$_{\rm maj}$} &
\colhead{FWHM$_{\rm min}$} &
\colhead{PA} &
\colhead{$T_{\rm SED}$} \\
\colhead{} &
\colhead{} &
\colhead{(J2000)} &
\colhead{(J2000)} &
\colhead{(mJy)} &
\colhead{($M_\odot$)} &
\colhead{(arcsec$^2$)} &
\colhead{(mJy)} &
\colhead{($M_\odot$)} &
\colhead{(mJy beam$^{-1}$)} &
\colhead{(\arcsec)} &
\colhead{(\arcsec)} &
\colhead{(deg)} &
\colhead{(K)}
}
\startdata
\hline
FIR1 & 1 & 06:46:15.83 & +00:06:28.4 & 54.90 & 68.25 & 6.16 & 42.70 & 53.08 & 6.97 & 2.35 & 0.90 & 84.1 & 21.5 \\
FIR1 & 2 & 06:46:19.89 & +00:05:07.6 & 5.70 & 7.09 & 1.44 & 4.70 & 5.84 & 3.71 & 0.72 & 0.60 & 29.0 & 15.0 \\
FIR1 & 3 & 06:46:15.65 & +00:06:29.2 & 4.60 & 5.72 & 1.00 & 8.80 & 10.94 & 1.29 & 2.20 & 1.07 & 40.9 & 21.7 \\
FIR1 & 4 & 06:46:18.81 & +00:05:08.4 & 3.70 & 4.60 & 1.44 & 3.50 & 4.35 & 2.02 & 0.86 & 0.70 & 100.1 & 15.3 \\
FIR1 & 5 & 06:46:21.71 & +00:04:47.8 & 2.50 & 3.11 & 1.12 & 2.80 & 3.48 & 1.54 & 0.90 & 0.70 & 88.0 & 15.6 \\
\enddata
\tablecomments{A full version of this table is available in machine-readable form. Core masses are calculated from the listed 1.3 mm fluxes assuming $T_{\rm dust}=20$ K, $D=4.5$ kpc, $\kappa_{1.3\,{\rm mm}}=0.899$ cm$^2$ g$^{-1}$, and metallicity $Z=0.5\,Z_\odot$. Fluxes and peak intensities are listed in mJy and mJy beam$^{-1}$, respectively. Gaussian-fit columns are reported only for robust cores with fitted Gaussian peak S/N $>5$; other entries are marked with ``-''. $T_{\rm SED}$ is sampled from the Herschel SED dust temperature map at the core position.}\label{table:core_cat}
\end{deluxetable}

\section{Artificial core insertion experiment}\label{sec:core_insert}

We use artificial core injection experiments to estimate both the flux recovery fraction, $f_{\rm flux}$, and the number recovery fraction, $f_{\rm num}$, for the CMF analysis. The injected masses are chosen to match the logarithmic mass-bin centers used for the CMF construction. These masses are converted to injected 1.3~mm flux densities using the fiducial mass--flux conversion adopted in the main analysis in \autoref{sec:core}. As noted by \citet{Kinman25}, the assumed core size can significantly affect the detection rate, because it changes the peak signal-to-noise ratio at fixed total flux. We therefore construct artificial sources whose morphologies and positions are tied empirically to the observed core population.

\begin{figure*}[ht!]
\epsscale{1.1}\plotone{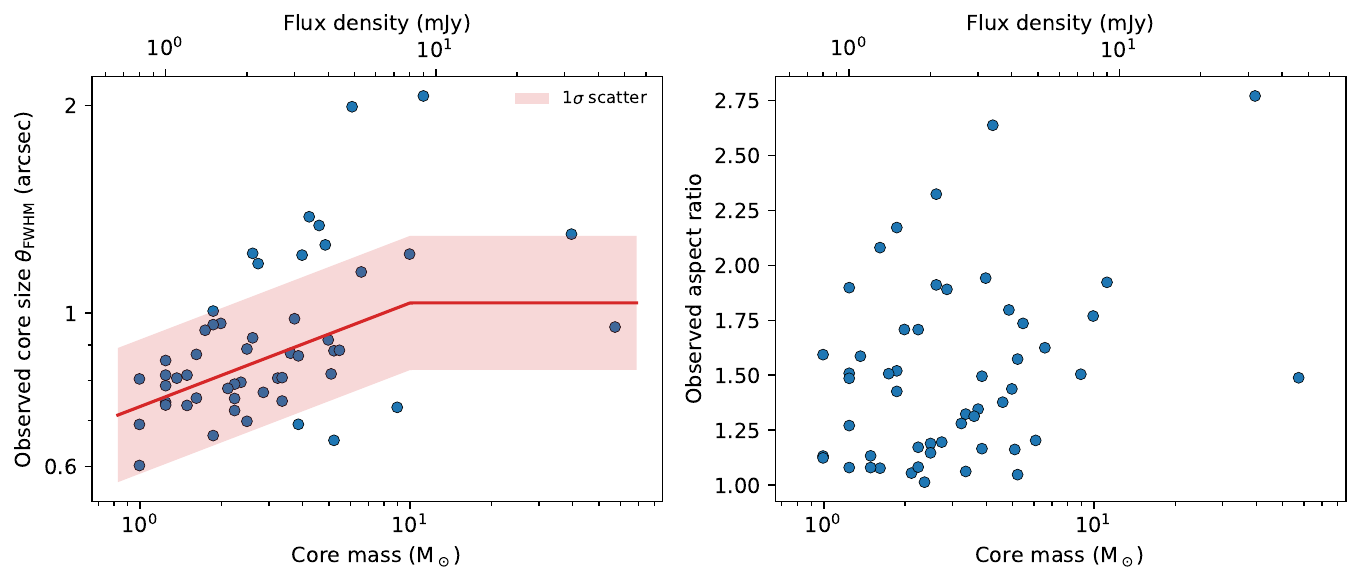}
\caption{Observed core sizes and aspect ratios as a function of core mass.
The left panel shows the observed core size, defined as the geometric-mean FWHM,
$\theta_{\rm FWHM}=(\theta_{\rm maj}\theta_{\rm min})^{1/2}$, without deconvolving
the ALMA synthesized beam. The red curve shows the empirical size model adopted for the core insertion experiments, obtained from a robust linear regression:
$\log_{10}(\theta_{\rm FWHM}/{\rm arcsec})=-0.135+0.150\log_{10}(M/M_\odot)$. The shaded region shows the 1$\sigma$ scatter, estimated as the standard
deviation of the residuals in $\log_{10}\theta_{\rm FWHM}$ around this
relation. The trend is held fixed above $10\,M_\odot$ since only a few cores are available. The right panel shows the observed aspect ratio,
$\theta_{\rm maj}/\theta_{\rm min}$. The upper axis gives the corresponding 1.3 mm flux density.
}\label{fig:core_property}
\end{figure*}

To characterize the observed core morphology, we use the Gaussian-fit results from the robust catalog, which provide more well-defined measurements of the source sizes, aspect ratios, and flux densities. We plot the observed core size (
$\theta_{\rm FWHM}=(\theta_{\rm maj}\theta_{\rm min})^{1/2}$) and aspect ratio ($A=\theta_{\rm maj}/\theta_{\rm min}$) against core masses in \autoref{fig:core_property}. We fit the observed sizes with a linear
relation in the $\log M$--$\log\theta_{\rm FWHM}$ plane, and obtain $\log_{10}(\theta_{\rm FWHM}/{\rm arcsec})=-0.135 + 0.150 \log_{10}(M/M_\odot)$. The scatter about this relation is
estimated from the standard deviation of the residuals in
$\log_{10}\theta_{\rm FWHM}$, giving $\sigma_{\log\theta}=0.097$ dex. For each
inserted core, we draw $\log\theta_{\rm FWHM}$ from this relation with a Gaussian
scatter of 0.097 dex, truncated at $\pm3\sigma$; the trend is held fixed above
$10\,M_\odot$ to avoid extrapolating beyond the sparsely sampled high-mass end.
The inserted size is also required to be no smaller than the synthesized beam in
the corresponding ALMA field. The aspect ratio is drawn independently from a
smoothed empirical distribution of the observed Gaussian aspect ratios.

The injection positions are also chosen empirically. For each ALMA field, we construct a two-dimensional KDE from the observed core positions, using a smoothing scale of 30\arcsec. Artificial-core positions are drawn from this KDE, so that the injections approximately follow the observed spatial distribution of dense cores and sample similar background and crowding conditions. For each flux density, we inject five artificial cores per field in each realization and repeat the experiment 1000 times.

\begin{figure*}[ht!]
\epsscale{1.1}\plotone{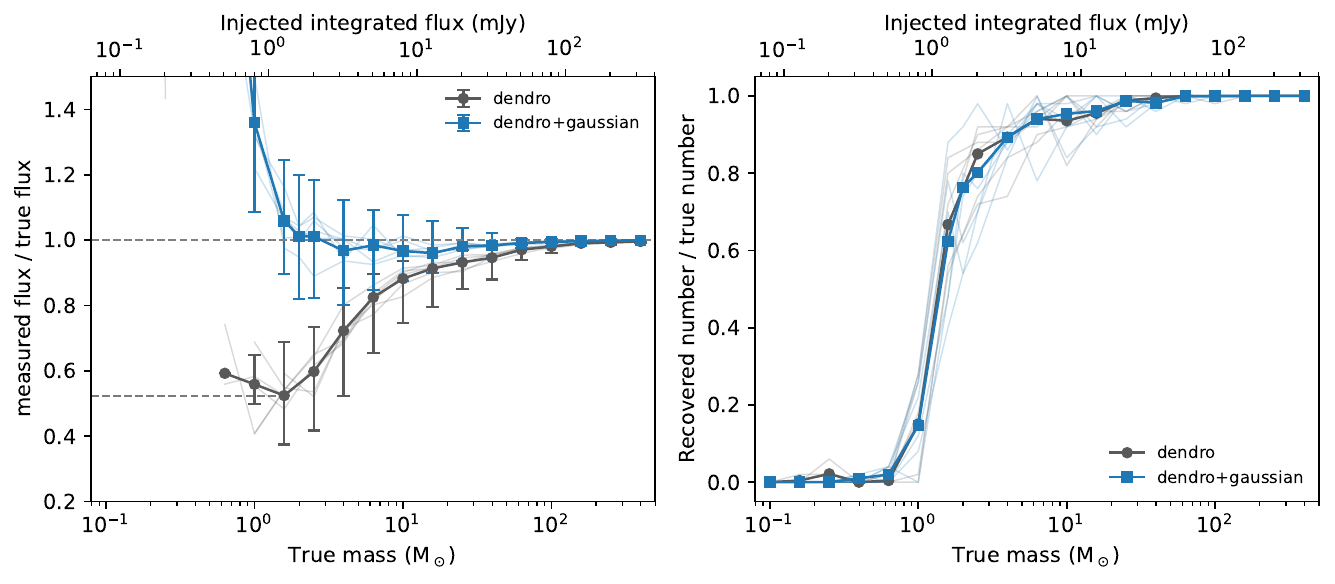}
\caption{Recovery curves from the artificial core injection experiments. Left: flux-recovery fraction, $f_{\rm flux}=F_{\rm meas}/F_{\rm true}$, as a function of injected true mass, with the corresponding injected 1.3~mm flux density shown on the top axis. Right: number recovery fraction, $f_{\rm num}=N_{\rm rec}/N_{\rm inj}$. Gray curves show the result for the dendrogram-only method, while blue curves show the dendrogram+Gaussian method. Thin curves show the results for individual ALMA fields, and the thick curves with points show the combined recovery functions weighted by core numbers. 
}\label{fig:recovery}
\end{figure*}

After each injection, we rerun the same core identification and flux measurement procedures used for the real data. An injected source is considered recovered if the peak pixel of a detected dendrogram leaf lies within two pixels of the injected position. In the pure dendrogram method, fluxes are measured by summing over the dendrogram-assigned pixels. In the dendrogram+Gaussian method, each dendrogram leaf is additionally fit with a Gaussian plus a constant background and is retained only if the fitted Gaussian component has a peak intensity greater than $5\sigma$; the fluxes are then measured from the fitted Gaussian models. For each injected flux or mass, we compute
$f_{\rm num}=N_{\rm rec}/N_{\rm inj}$ and
$f_{\rm flux}=F_{\rm meas}/F_{\rm true}$,
where $f_{\rm flux}$ is evaluated only for detected sources. The resulting recovery curves are shown in \autoref{fig:recovery}. Thin curves show the recovery functions measured separately for individual fields, while the thick curves show the core-number-weighted average over all fields. For the CMF correction, however, we treat each field independently and apply the recovery fractions appropriate for that field before combining the corrected core counts.

For the number recovery, both algorithms show similar trends. The pure dendrogram method has a higher $f_{\rm num}$ at $M\sim1$--$5~\msun$, as expected because the dendrogram+Gaussian catalog imposes an additional Gaussian peak-S/N criterion. {Note that the completeness derived here applies to the operationally defined
population of compact cores satisfying our intensity, contrast,
and area criteria. It quantifies the recovery of this population
against observational effects such as finite sensitivity, noise,
structured backgrounds, and source blending. Relatively diffuse or
extended cores that do not satisfy the adopted criteria may remain
undetected and are not represented by $f_{\rm num}$. Assessing such a
potentially distinct population would require synthetic observations
of numerical simulations \citep[e.g.,][]{Hsu23}, and is beyond the scope
of this work.}

The dendrogram+Gaussian method gives $f_{\rm flux}$ values close to unity over most of the fitted mass range, indicating stable flux recovery. In contrast, the pure dendrogram method tends to underestimate the integrated flux at low masses because of its isophotal pixel assignment. We therefore apply the flux-recovery correction only to the pure dendrogram catalog, while both catalogs are corrected for number completeness using their corresponding $f_{\rm num}$ curves.

\begin{figure*}
\centering
\includegraphics[width=0.5\textwidth]{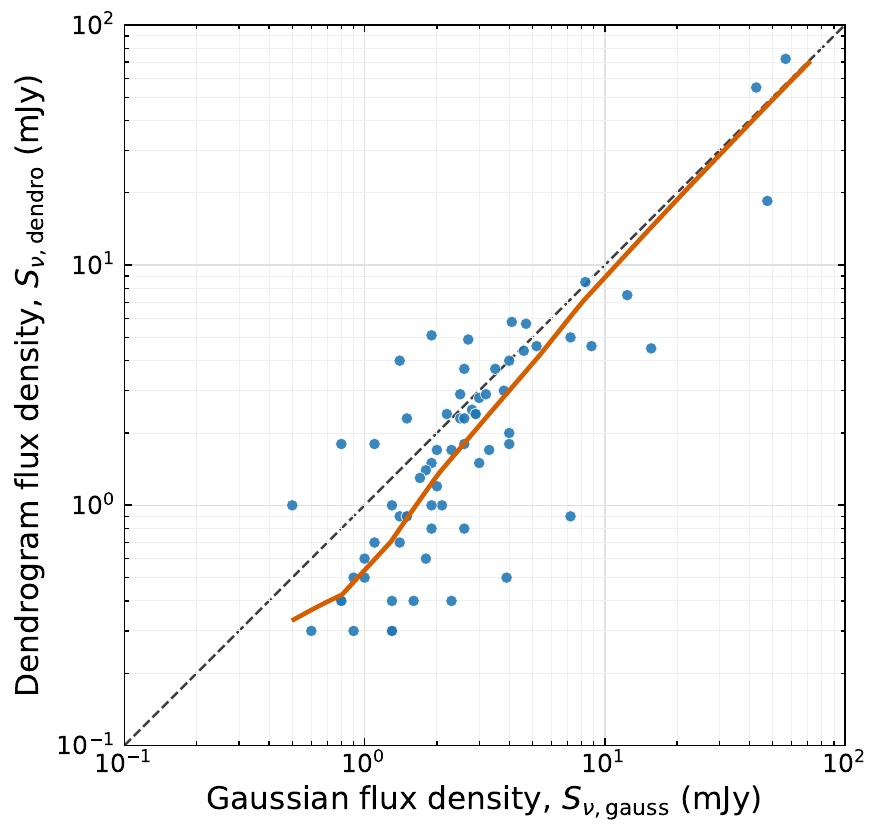}
\caption{Comparison between the dendrogram-integrated and Gaussian-fit
fluxes for the observed cores. The dashed line marks the one-to-one relation.
Colored curves show the expected relation
$S_{\nu,\rm dendro}=f_{\rm flux}S_{\nu,\rm gauss}$ by treating the Gaussian-fit flux as the reference source flux
and applying the corresponding flux recovery function derived from the
artificial core experiments.}
\label{fig:flux_compare}
\end{figure*}

{
\autoref{fig:flux_compare} compares the dendrogram-integrated and
Gaussian-fit fluxes for the observed cores. Although there is
substantial source-to-source scatter, most robust cores lie below the
one-to-one relation, suggesting that exclusion of low-intensity source
emission outside the leaf boundaries tends to dominate over the
additional background included within the apertures. This may partly
reflect that most cores are not located on strongly elevated
backgrounds. The curve shows the expected dendrogram fluxes obtained by treating the Gaussian-fit fluxes as estimates of the intrinsic core fluxes and multiplying them by the field-dependent $f_{\rm flux}$ functions derived from the artificial-core experiments. Their broad agreement with the observed distribution indicates that the insertion experiments capture the average flux bias of the dendrogram measurements.
The remaining scatter likely reflects differences in core morphology,
background structure, blending, and departures from the adopted
Gaussian-plus-constant model. Thus, $f_{\rm flux}$ should be interpreted
as a first-order, population-averaged correction for the net dendrogram
flux bias, rather than an exact source-by-source correction.
}

The $f_{\rm flux}$ curves drop to around 0.5 at $M\sim1$--$2~\msun$, but show an apparent upturn toward lower masses. This behavior reflects a selection effect near the detection limit. Around $1~\msun$, the peak S/N is close to 5, comparable to the detection thresholds of both algorithms. Lower mass injected cores therefore can be detected only in favorable cases, such as when they lie on top of a bright background. For the dendrogram method, the measured flux can then include a contribution from this background, artificially increasing the flux recovery fraction. This effect is less severe for the dendrogram+Gaussian method because the fit includes a local background term, although in some cases the Gaussian component may still absorb part of the structured hierarchical background emission. For the flux recovery correction, we therefore hold $f_{\rm flux}$ fixed at its minimum value toward lower masses, avoiding a non-monotonic mapping between observed flux and true mass. This treatment only affects masses below the adopted fitting threshold, the vast majority of which will not enter the high-mass CMF fit.

\section{Comparison of CMFs in different Galactic environments} \label{sec:cmf_comp}

A sequence of ALMA studies has measured CMFs across a broad range of Galactic star-forming environments using a largely uniform analysis framework \citep{Cheng18,Liu18,ONeill21,Kinman25}.
Paper I studied the massive protocluster
G286.21+0.17 \citep{Cheng18}; Paper II analyzed 30 clumps in
seven infrared dark clouds (IRDCs) \citep{Liu18}; Paper III
examined 28 high-mass-surface-density protocluster clumps
observed by ALMAGAL \citep{ONeill21}; and Paper IV extended
the sample to the Galactic center clouds: the Brick, Sgr~C, and
Sgr~B2-DS \citep{Kinman25}. These studies use the same basic
analysis framework: cores are identified as \textit{astrodendro}
leaves using a minimum intensity of $4\sigma$, a minimum
significance of $1\sigma$, and a minimum area of half the
synthesized beam; core masses are derived from millimeter
continuum emission using a fiducial dust temperature of 20~K;
and flux and number recovery are quantified through
artificial-core insertion experiments. Paper IV further updated
the insertion procedure to account for realistic, mass-dependent
core sizes and applied this revised correction framework to the
earlier samples. Our Sh2-284 analysis adopts the same dendrogram
criteria and an analogous recovery analysis, although the
gas-to-dust ratio is adjusted for its lower metallicity. This
series therefore enables a more controlled comparison between
environments.

\begin{figure*}[ht!]
\epsscale{1.1}\plotone{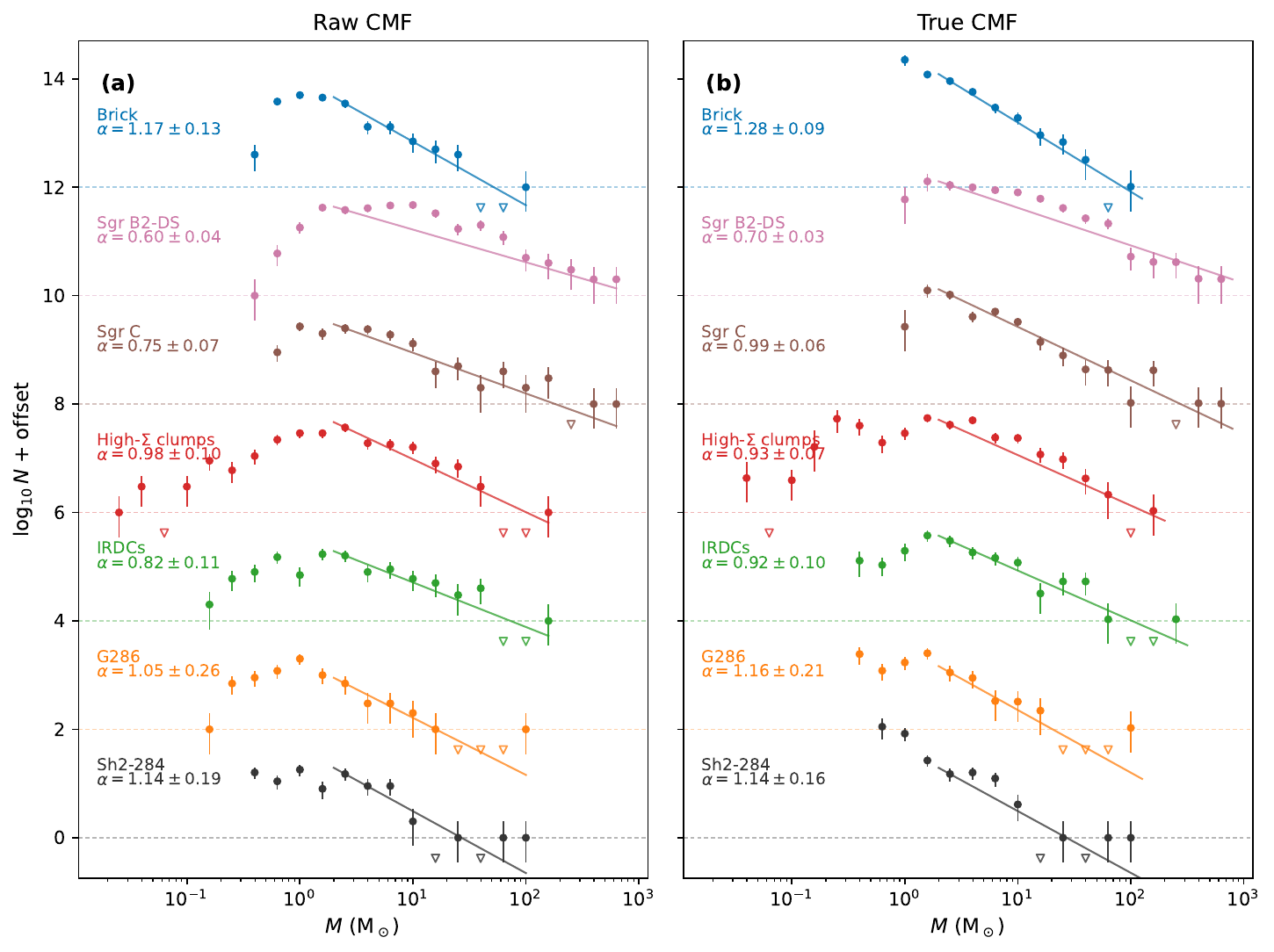}
\caption{Comparison of the CMFs in Sh2-284 and the Galactic
environments studied by \citet{Cheng18}, \citet{Liu18},
\citet{ONeill21}, and \citet{Kinman25}. Panel (a) shows the raw
CMFs, while panel (b) shows the true CMFs, in which masses are corrected for flux recovery
and the binned counts are corrected for number recovery. Dashed horizontal lines
mark the level corresponding to one core per bin for each
offset sequence, and open downward triangles mark zero-count
bins. Solid lines show power-law fits obtained uniformly with
the MLE method for panel (a) and binned MLE method for panel (b),
both starting from the bin centered at $2.51~M_\odot$ with a lower
edge of $1.995~M_\odot$. 
}\label{fig:cmf_comp}
\end{figure*}

\autoref{fig:cmf_comp} compares Sh2-284 with the
regions studied in this series (see also summary in \citet{Kinman25}). For Sh2-284, we use the full catalog, which most closely matches the
catalog-construction method used for the comparison samples.
All CMFs are constructed using the same 0.2-dex mass bins on
an aligned logarithmic grid. We show both the raw CMFs and
the true CMFs, in which the core
masses have been corrected for flux recovery and the binned
counts have been corrected for number recovery. To avoid
mixing slopes obtained using different fitting methods and mass
ranges, we re-fit the raw CMFs with the MLE method and the true CMFs with the binned MLE method of \citet{Virkar14}, all starting from 2~\msun. The
reported slopes can therefore differ slightly from those quoted
in the original papers. Note that the fit range is different from the fiducial Sh2-284 analysis in \autoref{sec:cmf}, which applies a threshold of $M>2.5~M_\odot$ to the individual masses or uses only complete bins above this threshold. The Sh2-284 slope reported in \autoref{fig:cmf_comp} should thus be regarded as the value obtained specifically for this homogeneous comparison.

The true CMF slopes span a broad range,
from $\alpha=0.70\pm0.03$ in Sgr~B2-DS to
$\alpha=1.28\pm0.09$ in the Brick. The Brick has the steepest,
approximately Salpeter-like CMF, while Sgr~B2-DS has a much
shallower distribution. Sgr~C, the IRDCs, and the high-$\Sigma$
protocluster clumps occupy intermediate or relatively shallow
ranges. Sh2-284, with $\alpha=1.14\pm0.16$ under the common
fitting prescription, lies close to G286
($\alpha=1.16\pm0.21$) and toward the steeper side of the
overall distribution. Thus, even when similar identification,
correction, binning, and fitting procedures are adopted, the
CMFs do not converge to a single universal high-mass slope.

\citet{Kinman25} suggested that the diversity of CMF slopes may reflect both environmental conditions and evolution. Regions with higher core mass surface densities may preferentially develop shallower CMFs, while continued accretion onto massive cores could flatten the CMF as star formation proceeds. However, the full sample does not follow a simple evolutionary sequence.
Systematic temperature variations and differences in angular resolution, spatial filtering, and the mixture of prestellar and protostellar cores remain important caveats.

Within these limitations, the location of Sh2-284 toward the
steeper part of the distribution is qualitatively consistent with
its relatively low density and less extreme star forming
environment. In particular, it indicates that moderately low
metallicity alone does not necessarily produce the shallow,
top heavy CMFs found in some dense and actively star forming
regions. Instead, the comparison likely suggests that the high-mass
CMF is regulated by a combination of cloud environment,
evolutionary state, and subsequent core growth.

\bibliographystyle{aasjournal}
\bibliography{refer}

\end{CJK}
\end{document}